\documentclass{aa}

\usepackage{graphicx}
\usepackage{epsfig}

\begin{document}
    
\title
{The statistical significance of the North-South asymmetry of solar
activity revisited}

\author
{M. Carbonell$^1$ \and J. Terradas$^2$ \and R. Oliver$^2$ \and J. L. Ballester$^2$}

\offprints{J. L. Ballester\\ \email{joseluis.ballester@uib.es}}

\institute{Departament de Matem\`atiques i Inform\`atica, Universitat de les Illes Balears, 
E-07122 Palma de Mallorca, Spain
\and
Departament de F\'{\i}sica, Universitat de les Illes Balears, 
E-07122 Palma de Mallorca, Spain \\
\email{marc.carbonell@uib.es; jaume.terradas@uib.es;
ramon.oliver@uib.es; joseluis.ballester@uib.es}}

\date{Received / Accepted}

\abstract 
{} 
{Many studies of the North-South asymmetry of solar activity and
its features have been performed.  However, most
of these studies do not consider whether or not the asymmetry of the time
series under consideration is statistically significant. If the asymmetry is statistically insignificant, any
study about its behavior is meaningless. Here, we discuss the difficulties found when trying to assess
the statistical significance of the North-South asymmetry (hereafter SSNSA) of the most
usually considered time series of solar activity.} 
{We distinguish between solar activity time series 
composed of integer or non-integer and dimensionless data, or composed 
of
non-integer and dimensional data. For each of these cases, we discuss 
the most suitable statistical tests which can be applied and highlight the 
difficulties in obtaining valid information about the statistical
significance of solar activity time series.} 
{Our results suggest that, apart from the need to apply suitable
statistical tests, other effects such as data binning, the
considered units and the need, in some tests, to consider groups of
data, substantially affect the determination of the statistical
significance of the asymmetry.} 
{The assessment of the statistical
significance of the N-S asymmetry of solar activity is difficult
and an absolute answer cannot be given, since many
different effects influence the results given by the statistical
tests.  The quantitative results about the statistical
significance of the N-S asymmetry of solar activity provided by
different authors, as well as studies of its behaviour, must be
considered with care because they depend on the chosen values of
different parameters or on the considered units.}
{}

\titlerunning{Statistical significance North-South Asymmetry} 
\authorrunning{Carbonell et al.} 
\maketitle

\section{Introduction}
The North-South (N-S) asymmetry of solar activity has been the subject
of many studies using different features of solar
activity.  Some of the most important features considered have been: the number of flares (Roy 1977; Garcia 1990;
Verma 1987, 1992; Li et al.  1998; Temmer et al.  2001), the flare
index (Kno\v{s}ka 1985; Ata\c{c} \& \"{O}zg\"{u}\c{c} 1996, 1998;
Joshi \& Joshi 2004), the sunspot number and sunspot areas (Newton \&
Milson 1955; Waldmeier 1957, 1971; Roy 1977; Swinson et al.  1986;
Vizoso \& Ballester 1990; Schlamminger 1991; Verma 1992; Yi 1992;
Carbonell et al.  1993; Verma 1993; Oliver \& Ballester 1994;
Pulkkinen et al.  1999; Li et al.  2000; Li et al.  2002; Vernova et
al.  2002; Temmer et al.  2002; Li et al.  2003; Knaack et al.  2004; 
Temmer et al. 2006),
the photospheric magnetic flux (Rabin et al.  1991; Knaack et al.
2004) and the number of solar active prominences (Joshi, 1995; Verma,
2000).

There is an important solar activity feature, the
photospheric magnetic flux, for which the behavior of the N-S
asymmetry has been also studied.  Howard (1974), using magnetic flux
data from Mt.  Wilson, analyzed the period between 1967 and 1973;
Rabin et al.  (1991) using magnetic flux data from Kitt Peak, studied
the period between 1975 and 1987; Knaack et al.  (2004) have used Kitt
Peak data about photospheric magnetic flux density to study the time
interval between 1975 and 2003; Song et al.  (2005) have used
Kitt Peak data, between 1978 and 2002.  However, of all these authors only
Song et al.  (2005) have estimated the statistical significance of the
North-South asymmetry (hereafter SSNSA) of the photospheric magnetic
flux time series under consideration, applying the
binomial distribution, and so most of the conclusions of the rest of
studies about the N-S asymmetry of photospheric magnetic flux have
been obtained by visual inspection of the plot of the
asymmetry versus time.  Taking into account the fundamental role of
the magnetic flux in solar activity, a quantitative assessment of its
N-S asymmetry should be of great interest.

Before studying the behavior of the N-S asymmetry, the most important
point, sometimes forgotten, is to assess the SSNSA of the time series under
consideration.  The most straightforward way to determine the SSNSA is
by means of the binomial distribution (Li et al.  1998; Li et al.
2003; Song et al.  2005).  However, other statistical tests have also
been used, for instance, Joshi (1995), Temmer et al.  (2001) and Joshi
\& Joshi (2004) have followed Leftus (1960) using a $\chi^2$-test to
assess the SSNSA of sunspot groups, H$\alpha$ flares and active
prominences; Temmer et al.  (2002; 2006) have used a paired Student's
test to study the SSNSA of the hemispheric sunspot number; Ata\c{c}
and \"{O}zg\"{u}\c{c} (1996) used a sign test, introduced by
Gleissberg (1947), to determine the SSNSA of the flare index, and
Vizoso and Ballester (1990) and Carbonell et al.  (1993) have used
Excess (Reid, 1968; Wilson, 1987) to obtain the SSNSA of sunspot
areas.  In all the cases mentioned before, the authors concluded
that, to a great extent, the N-S asymmetry of the considered time
series was statistically significant.

However, a blind application of the above statistical tests to any
considered solar activity time series can lead to misleading results.
The commonly considered solar activity time series have two
different forms: (a) composed of integer or non-integer
data; (b) composed of dimensional or dimensionless data.  The first
characteristic, integer or non-integer data, is relevant for the statistical tests 
that can be applied.  In the case of integer
data records any statistical test can be applied, but in the case of
non-integer data one needs to choose carefully what tests can be used.
Related to the second characteristic, dimensional or dimensionless
data, this raises two interesting problems when dimensional
data are considered: (a) How does a change of units affect the statistical
significance?; (b) Is it possible to find a statistical test whose
results are independent of the considered units?

Here, we search for quantitative conclusions about the SSNSA of the
most often considered solar activity time series.  To this end, we
analyse the results obtained after the application of different
statistical tests to solar activity time series made of integer and
dimensionless data, and the effects on the SSNSA induced by data
binning and point out the problems encountered when
trying to assess the SSNSA of non-integer and dimensional solar
activity time series. We perform these analysis since many conclusions about the behaviour of the N-S
asymmetry of solar activity have been extracted without a proper
quantitative evaluation of its statistical significance, or without
considering whether a definitive answer about the SSNSA can be obtained.

\section{Data and methods}

\subsection{Data}
The different solar activity time series analyzed in our study are:
\begin{itemize}
\item The daily north and south international sunspot number (1992 -
2004), which can be downloaded from
http://www.ngdc.noaa.gov/stp/SOLAR/ \\ ftpsunspotnumber.html, with $4718$ data
points.
\item The daily north and south hemispheric sunspot numbers (1945 - 2004), 
 compiled by Temmer et al. (2006),  
 which can be downloaded from
 http://cdsweb.u-strasbg.fr/cgi-bin/qcat?J/A+A/447/735, with 21915
 data points.
\item The yearly north and south number of active prominences (1957 -
 1998), obtained from Verma (2000), with $42$ data points.
\item The daily north and south X-ray flares (1975 - 2006), detected by the GOES
satellites, which can be downloaded from http://www.ngdc.noaa.gov/stp/SOLAR/
\\ ftpsolarflares.html,
with $11323$ data points.
\item The daily number of north and south solar flare index (1976 -
2004) compiled in the Kandilli Observatory, which can be downloaded from
http://www.ngdc.noaa.gov/stp/SOLAR/ \\ ftpsolarflares.html, with 10227 data
points.
\item The monthly north and south sunspot area between 
May 1874 and January 2004 compiled by D. Hathaway, and which can be
downloaded from http://science.msfc.nasa.gov/ssl/pad/solar/ \\ greenwch.htm, with 
$1557$ data points.
\item The north and south NSO/Kitt Peak averaged photospheric magnetic
flux density corresponding to Carrington rotations 1625 - 2007
(February 1975
- August 2003), with $383$ data points.
\item The north and south total magnetic flux from Mt. Wilson
Observatory corresponding to Carrington rotations
1511 - 2013 (September 5, 1966 - August 26, 2004), with $502$ data points.

\end{itemize} 

\subsection{Preliminary considerations}

Usually, the application of a statistical test is based on the
hypothesis of independence of experiments.
However, when we consider the N-S asymmetry of solar activity, the
data records of many of the considered time series are not independent.  For instance, if we consider daily sunspot areas,
magnetic flux, solar prominences, sunspot number, etc. the values of these quantities today are not independent
of
the values yesterday, in particular, daily sunspot areas are 
correlated with a typical correlation time of about $7$ days (Oliver
\& Ballester, 1995). 
Then, strictly speaking, solar activity does not satisfy the above
hypothesis, and, for this reason, the application of statistical tests
to determine the SSNSA is not appropriate.
Another consideration of interest is related to the characteristics of
the studied solar activity time series which can be
classified as composed of integer and dimensionless data, such as the
International sunspot number, the number of X-ray flares and the
number of solar active prominences, or composed of non-integer and
dimensional data, such as the $H_{\alpha}$ flare index, sunspot areas,
Mount Wilson total magnetic flux and Kitt Peak averaged magnetic flux
density.  This second consideration is of paramount importance since
it constrains the statistical tests that can be applied to the
above time series. The third consideration is that in all the
studies about the N-S asymmetry we only consider the solar activity
corresponding to the visible hemisphere.

 \begin{table*}
  \small
  \centering
  \begin{tabular}{cccccccc}
  \hline
  \hline
   ND &  \ \ \ \ \ \ \ \ & $\chi^2$ & \ \ \ \ \ \ \ \ 
  & E & 
  \ \ \ \ \ \ \ \ & P(d) & \ \ \ \ \ \ \ \ \\
  \hline
  \hline
                       7817 (93.6 \%) & $$ & 7749 (92.8 \%)  & $$ &
		       7693 (92.1 \%) & $$ & 7728 (92.6 \%)\\
  \hline
                       165 (1.9 \%)  & $$ & 163 (1.9 \%) & $$ &
		       170 (2.0 \%) & $$ & 158 (1.9 \%) \\
  \hline
                       91 (1.1 \%) & $$ & 70 (0.8 \%)   &  $$ &
		       180 (2.1 \%)  &  $$ & 85 (1.0 \%)\\
 \hline 
                       272 (3.2 \%) & $$ & 363 (4.3 \%) & $$ & 302
		       (3.6 \%)  &  $$ & 374 (4.5 \%)\\
 \hline
  \label{p}
  \end{tabular}
  \begin{center}
  \caption{Surface expressed in $dm^2$.  The labels of the columns,
  from left to right, correspond to the normal distribution
  approximation to the binomial distribution, to the Pearson's
  chi-square test, to the Excess, and to the binomial distribution.
  The rows, from top to bottom, correspond to a highly statistically
  significant result, a statistically significant result, a marginally
  significant result, and a statistically insignificant result.  The
  numbers shown in each column correspond to the number of events and
  its corresponding percentage with respect to the data records of the
  considered time series.}
  \end{center}
  \end{table*}

\subsection{Statistical tests for dimensionless time series}
\label{se}
To assess the SSNSA of solar activity time series, we
distinguish between integer and dimensionless time series,
non-integer and  dimensionless time series, and
non-integer and dimensional time series. In the case of integer and dimensionless time
series we have applied the following statistical tests:
 
 \begin{enumerate}
    
\item {Binomial distribution}: When the data records of the considered
time series are integer numbers, the binomial formula (Larson,
1982) can be used to
compute the probability of obtaining any particular
distribution of $n$ objects into two classes.  In our case, these two
classes correspond to North and South hemispheres ($N$ and $S$).  The binomial
formula is given by $$P(r)=\frac{n!}{r!(n-r)!}p^r(1-p)^{n-r},$$
 where $n$ is the number of objects in both classes, $r$ is the number
 of objects in a particular class, and $p$ is the probability
 associated with that particular class, in our case $0.5$.  An
 alternative way (Wilson, 1987) is to describe the probabilities using the
 difference $d=|r-(n-r)|$ with $P(d)=2\cdot P(r)$ for all $P(r)$
 except when $d=0$, and $P(\geq d)$ is given by $$P(\geq
 d)=\mathop{\sum}\limits_{i=d}^{n}P(i).$$
In general, when $P(\geq d)>10\%$ this implies a statistically
insignificant result; when $5\%<P(\geq d)<10\%$ it is marginally
significant; when $1\%<P(\geq d)<5\%$ we have a statistically
significant result and when $P(\geq d)<1\%$ the result is highly
significant.
 
 \item {Excess}: 
 An approximation to compute $P(\geq d)$ is the Excess (Reid, 1968; Wilson, 1987), which is a measure 
 proportional to the
 uncertainty, computed as $d(n/2)^{-1/2}$, where $d=|N - S|$ is the
 positive difference between the two hemispheres and $n$ is the total
 quantity corresponding to both hemispheres. The equivalence of Excess and
 $P(\geq d)$ is given by: Excess $<$ 2 with $P(\geq d)>10\%$;
 2 $<$ Excess $<$ 3 with $5\%<P(\geq d)<10\%$; 3 $<$ Excess $<$ 4 with
 $1\%<P(\geq d)<5\%$ and 4 $<$ Excess with $P(\geq d)<1\%$.
 \item {Normal approximation to the Binomial distribution}: Another
  way to study the statistical significance of an asymmetry time
  series is to consider the normal
  distribution as an approximation of the binomial distribution
  (Larson, 1982).  In
  this case, the normal distribution has the same mean $\mu = np$ and
  standard deviation $\sigma=\sqrt{np(1-p)}$ as the binomial
  distribution.  Then, we have considered the standard normal
  distribution with $p = 0.5$
  $$
  Z=\frac{X-\mu}{\sigma}=\frac{N-S}{\sqrt{(N+S)}}.
  $$
 To test the SSNSA of our time
 series, we have counted the
 number of events in which the statistic $Z$ lies in different
 intervals $0\leq Z\leq 0.01$ or $0.99\leq Z\leq 1$, which indicate
 that the asymmetry is highly significant.
 
 \item {Pearson's chi-square test ($\chi^2$)}: For this goodness of fit
 computation (Larson, 1982), the data are divided into $k$ classes and the test
 statistic is defined as $$\chi^2 = \mathop{\sum}_{i=1}^{k} \frac{(O_i
 - E_i)^2}{E_i},$$ where $O_i$ represents the observed frequency and $E_i$
 the expected (theoretical) frequency, asserted by the null hypothesis
 $H_0$.  In our case, $H_0$ is that the observed data follow a
 binomial distribution with $p=1/2$.  So our statistic is:
 $$
 \chi^2=\mathop{\sum}_{i=1}^{2} \frac{(N_i - np_i)^2}{np_i} = \frac{(N -
 S)^2}{N+S}.
 $$
 The obtained test statistic is then compared with a chi-square
 distribution $\chi^2_{1,\alpha}$ with one degree of freedom, since we
 only consider two classes.  The significance level $\alpha$ is taken
 in such a way that when $\chi^2 \leq 0.01$ (highly significant
 asymmetry), $0.01 < \chi^2 \leq 0.05$ (significant asymmetry), $0.05
 < \chi^2 \leq 0.1$ (marginal asymmetry); $0.1 < \chi^2 \leq 1$
 (insignificant asymmetry).
 
 \end{enumerate}
 Probably, the most suitable test to be applied in order to determine
 the SSNSA is the binomial distribution test.  However, this test can
 only be applied when the data elements of the considered time series
 are integers, which constrains its applicability to time series of
 solar activity made of the number of events happening in each solar
 hemisphere.  On the other hand, this test has to be used properly, as
 we now illustrate by considering two synthetic time series
 corresponding to the North and South hemispheres, respectively.  Let
 us assume that we construct two time series composed of $105$ data
 points taken, for instance, at consecutive Carrington rotations.  In
 the Northern hemisphere, the first $63$ data points have a value of
 $6$ while the remaining $42$ have a value of $4$; in the Southern
 hemisphere the first $63$ data points have a value of $4$ and the
 remaining $42$ have a value of $6$.  Song et al.  (2005) consider
 that for $63$ rotations the northern hemisphere has been dominant and
 they take this quantity as $d$ while $n$ has the value of $105$.  In
 this way the result of the test is independent of the values of
 quantities in the Northern and Southern hemispheres.  Computing
 $P(d)$ gives a value of $0.0252$ ($2.5 \%$) which suggests a strong
 statistical significance.  However, a proper application of the
 binomial distribution is based on determining the SSNSA of each pair
 of values (North and South) corresponding to the same Carrington
 rotation, evaluating, finally, the number of pairs whose SSNSA is
 highly significant.  Then, after applying this procedure to the above
 considered time series we obtain that $P(d)$ is greater than $10 \%$
 in all of the $105$ cases.  This means that the SSNSA of these
 synthetic data points  is insignificant in all the Carrington rotations as
 should be expected from the small difference between the Northern and
 Southern hemisphere values.
 
In the case of non-integer and dimensionless time series, only the
tests Excess, Normal approximation to the Binomial distribution, and
Pearson's chi-square test can be applied.

\subsection{Statistical tests for non-integer and dimensional time series}
 
When the data are dimensional another problem appears related
to the considered units.  In solar activity time series, the
accuracy of the non-integer and dimensional data is determined by the
measurement process, so the data are truncated after some decimal
places.  One way to obtain integer data is to modify the units
of the considered time series, and to apply the statistical tests
described in the section \ref{se}.  Then, an interesting experiment is
to consider what happens to the statistical significance when the units
are modified.  A simple way to do this is to generate two synthetic
time series, corresponding to Northern and Southern hemispheres, made of
non-integer and dimensional data.  We have chosen two time series
composed of data records representing surfaces expressed in square
meters up to two decimal places.  These non-integer data can be
transformed to integer data by multiplying by $10^2$ to obtain surfaces
in $dm^2$, or by multiplying by $10^4$ to obtain surfaces in $cm^2$.  To
the resulting time series we have applied the tests of the previous
section and the results are shown in Tables 1-2 which point out that,
when going from $dm^2$ to $cm^2$, the SSNSA changes and increases.
These results suggest that transforming non-integer dimensional
data to integer data by changing the units modifies the statistical
significance. These results point out a difficult problem
because when dealing with dimensional time series the SSNSA will depend
on the considered units, at least when the above tests are used. Our problem now is to find a statistical test applicable to
non-integer and dimensional time series, and whose results are
independent of the considered units.

  \begin{table*}
  \small
  \centering
  \begin{tabular}{cccccccc}
  \hline
  \hline
   ND &  \ \ \ \ \ \ \ \ & $\chi^2$ & \ \ \ \ \ \ \ \ 
  & E & 
  \ \ \ \ \ \ \ \ & P(d) & \ \ \ \ \ \ \ \ \\
  \hline
  \hline
                       8296 (99.4 \%)  & $$ & 8293 (99.3 \%)  & $$ &
		       8288 (99.3 \%) & $$ & 8293 (99.3 \%)\\
  \hline
                       16 (0.2 \%)  & $$ & 17 (0.2 \%) & $$ &
		       14 (0.1 \%) & $$ & 17 (0.2 \%)\\
  \hline
                       8 (0.1 \%) & $$ & 2 (0.02 \%) &  $$ &
		       15 (0.1 \%) &  $$ & 2 (0.02 \%) \\
 \hline 
                        25 (0.3 \%) & $$ & 33 (0.4 \%) & $$ & 28
			(0.3 \%)  &  $$  & 33 (0.4 \%)\\
 \hline
 \label{table:parameters7}
  \end{tabular}
   \begin{center}
  \caption{Surface expressed in $cm^2$. Columns, rows and numbers in
  the rows have the same meaning as in Table 1.}
  \end{center}
  \end{table*}
 
  \subsection{The Student's t-test} \label{se1}
  
  A suitable test satisfying the above conditions is the
  paired Student's t-test (Larson, 1982).  The characteristic statistic $\hat t$ is
  expressed as
  $$
  \hat t = \frac {\sum D_{i}/n}{\sqrt \frac{\sum D^2_{i}-(\sum
  D_{i})^2/n}{n(n-1)}},$$
  where $D_{i}$ is the difference of paired values and $n$ represents
  the number of elements that we have in each of the groups in which
  we split the corresponding time series, $n - 1$ being the number of
  degrees of freedom.  Once the statistic $\hat t$ has been
  calculated, and taking into account the degrees of freedom, it is
  compared with $\hat t_{n-1}, _{\alpha}$ given in statistical tables
  on a previously chosen probability $\alpha$.  The important feature
  of this test is that when the units are changed, the same factor appears
  in both the numerator and denominator, and the value of the $\hat
  t$-statistic is not modified.  This can be ckecked using the
  surface time series, in $m^2$, $dm^2$, and $cm^2$, considered in the
  previous section. Then, taking $n = 30$ we obtain the same
  results for the three time series i.  e. the value of the
  statistic $\hat t$ is: $> 0.99$ for $32.05\%$ of the groups; between
  $0.95$ and $0.99$ for $18.82\%$ of the groups; between $0.9$ and
  $0.95$ for $14.70\%$ of the groups; and smaller than $0.9$ for
  $34.11\%$ of the groups.  Taking this feature into account, the Student's test can be very useful for the
  determination of the SSNSA of non-integer and dimensional time
  series.

\section{Results}

\subsection{Results for integer and dimensionless time series}
The statistical tests mentioned in section~\ref{se} have been
applied to the time series corresponding to the daily international
sunspot number, the yearly number of solar active prominences and the
daily number of X-ray flares, and the obtained results are displayed
in Tables~3-5.  In
Table~3, the SSNSA corresponding to the daily
international sunspot number is shown and a good agreement between
the results obtained with the different tests can be seen.
Also, the results show that the N-S asymmetry of the time series is
only highly significant or statistically significant on about 50 - 60
\% of the considered days.  In Table~4, the
results obtained for the yearly number of solar active prominences are
shown.  These results show a striking agreement in the statistical
significance between the different tests, and suggest that the N-S
asymmetry is highly significant in about 85 \% of the considered
years.  In Table~5, the results for the
daily number of solar X-ray flares are shown.  These results indicate
a very low statistical significance (between 3 - 7 \% of the days) for the N-S
asymmetry for this time series, which suggests that this solar
activity feature displays no significant N-S asymmetry.

  \begin{table*} 
  \small
  \centering
  \begin{tabular}{cccccccc}
  \hline
  \hline
   ND &  \ \ \ \ \ \ \ \ & $\chi^2$ & \ \ \ \ \ \ \ \ 
  & E & 
  \ \ \ \ \ \ \ \ & P(d) & \ \ \ \ \ \ \ \ \\
  \hline
  \hline
                       2608 (59.2 \%)  & $$ & 2463 (55.9 \%)  & $$ &
		       2307 (52.3 \%) & $$ & 2398 (54.4 \%) \\
  \hline
                       451 (10.2 \%)  & $$ & 410 (9.3 \%) & $$ &
		       449 (10.2 \%) & $$ & 394 (8.9 \%)\\
  \hline
                       220 (5.0 \%) & $$ & 186 (4.2 \%)  &  $$ & 440 (10 \%) 
		        &  $$ & 198 (4.5 \%) \\
 \hline 
                       1128 (25.6 \%)  & $$ & 1348 (30.6 \%) & $$ & 1211 (27.5 \%)  &  $$ & 1417
		       (32.2 \%) \\
 \hline
  \label{table:parameters1}
  \end{tabular}
  \begin{center}
  \caption{Daily international sunspot number.
  Columns, rows and numbers in
  the rows have the same meaning as in Table 1.}
  \end{center}
  \end{table*}

   \begin{table*} 
  \small
  \centering
  \begin{tabular}{cccccccc}
  \hline
  \hline
   ND &  \ \ \ \ \ \ \ \ & $\chi^2$ & \ \ \ \ \ \ \ \ 
  & E & 
  \ \ \ \ \ \ \ \ & P(d) & \ \ \ \ \ \ \ \ \\
  \hline
  \hline
                       36 (85.7 \%)  & $$ & 36 (85.7 \%)  & $$ &
		       35 (83.3 \%) & $$ & 35 (83.3 \%) \\
\hline
                       0 (0 \%)  & $$ & 1 (2.4 \%) & $$ &
		       1 (2.4 \%) & $$ & 1 (2.4 \%)\\
  \hline
                       2 (4.8 \%)  & $$ & 0 (0 \%) & $$ &
		       2 (4.8 \%) & $$ & 0 (0 \%)\\
  \hline
                       4 (9.5 \%) & $$ & 6 (14.3 \%)  &  $$ & 4 (9.5 \%) 
		        &  $$ & 6 (14.3 \%)\\
 \hline
  \label{table:parameters2}
  \end{tabular}
   \begin{center}
  \caption{Yearly number of active solar prominences. 
  Columns, rows and numbers in
  the rows have the same meaning as in Table 1.}
  \end{center}
  \end{table*}
  
  \begin{table*}
  \small
  \centering
  \begin{tabular}{cccccccc}
  \hline
  \hline
   ND &  \ \ \ \ \ \ \ \ & $\chi^2$ & \ \ \ \ \ \ \ \ 
  & E & 
  \ \ \ \ \ \ \ \ & P(d) & \ \ \ \ \ \ \ \ \\
  \hline
  \hline
                       646 (7.5 \%)  & $$ & 386 (4.5 \%)  & $$ &
		       277 (3.2 \%) & $$ & 277 (3.2 \%) \\
  \hline
                       1250 (14.6 \%)  & $$ & 834 (9.7 \%) & $$ &
		       689 (8.0 \%) & $$ & 439 (5.1 \%)\\
  \hline
                       1100 (12.8 \%) & $$ & 676 (7.9 \%)   &  $$ &
		       1703 (19.9 \%) 
		        &  $$ & 394 (4.6 \%)\\
 \hline 
                        5568 (65 \%) & $$ & 6668 (77.8 \%) & $$ & 5895
			(68.9 \%)  &  $$ & 7454
		       (87.0 \%) \\
 \hline
  \label{table:parameters3}
  \end{tabular}
  \begin{center}
  \caption{Daily number of X-ray flares.
  Columns, rows and numbers in
  the rows have the same meaning as in Table 1.}
  \end{center}
  \end{table*}

\subsubsection{Effect of data binning}

\begin{figure}[ht]
       \centering{
      \epsfig{file=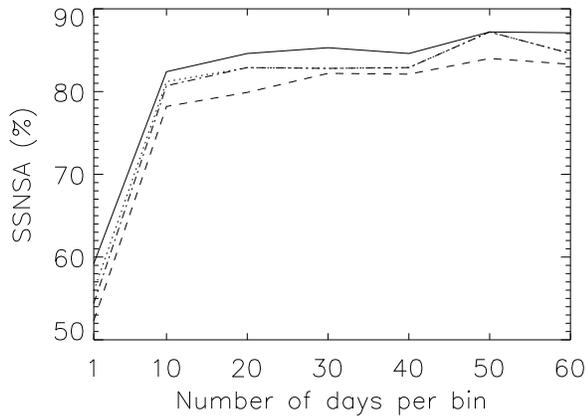,width = 9cm}}
                  \caption{SSNSA of the international sunspot number
		  time series
		  versus the number of days per bin for the binomial
		  distribution (dash - dot line); Excess (dashed line);
		  normal approximation to the binomial distribution
		  (solid line); and Pearson's chi-square test (dotted line).}             
   \label{sig1} 
          \end{figure}

\begin{figure}[ht]
       \centering{
      \epsfig{file=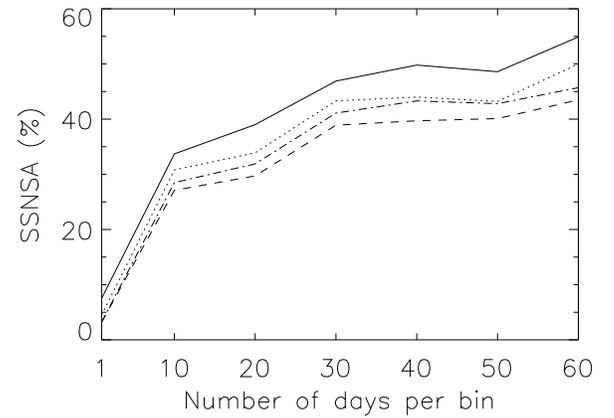, width = 9cm}}
                  \caption{SSNSA of the number of X-ray flare time series
		  versus the number of days per bin for the binomial
		  distribution (dash - dot line); Excess (dashed line);
		  normal approximation to the binomial distribution
		  (solid line); and Pearson's chi-square test (dotted line).}             
   \label{sig2} 
          \end{figure}

We study the effect of binning data on the statistical significance.
We have binned the time series corresponding to the international
sunspot number and X-ray flare in bins of $10$, $20$, $30$, $40$, $50$
and $60$ days.  The same statistical tests as in previous section have
been applied to these new time series and in
Figs~\ref{sig1}-\ref{sig2} the SSNSA versus the number of days per bin
is plotted.  Binning data modifies the statistical significance, and
the general trend is an increase of the SSNSA. If one starts from two
north and south daily time series and computes the SSNSA, the obtained
results would be different from those obtained by binning the times
series for Carrington rotation and computing, again, the statistical
significance.  In order to compute a meaningful statistical asymmetry,
what is the appropiate data binning, if any, to be considered?

\subsection{Results for non-integer and dimensionless time series}

The only considered solar activity time series whose data records are
non-integer and dimensionless is that of hemispheric sunspot numbers.
To this time series we have applied the tests given in
section~\ref{se1}.  The results are shown in Table 6, which shows that
there is a strong agreement between the statistical significances
obtained from the different tests.  Furthermore, the results indicate
that the N-S asymmetry of hemispheric sunspot numbers is highly
significant in about 60 \% of the considered days.

 \begin{table*}
  \small
  \centering
  \begin{tabular}{cccccccc}
  \hline
  \hline
   ND &  \ \ \ \ \ \ \ \ & $\chi^2$ & \ \ \ \ \ \ \ \ 
  & E & 
  \ \ \ \ \ \ \ \ & P(d) & \ \ \ \ \ \ \ \ \\
  \hline
  \hline
                       13025 (64.83 \%)  & $$ & 12348 (61.46 \%)  & $$ &
		       11547 (57.46 \%) & $$ & - \\
  \hline
                       1934 (9.62 \%)  & $$ & 1694 (8.43 \%) & $$ &
		       2061 (10.25 \%) & $$ & -\\
  \hline
                       1076 (5.35 \%) & $$ & 917 (4.56 \%)   &  $$ &
		       2042 (10.16 \%) 
		        &  $$ & -\\
 \hline 
                        4055 (20.18 \%) & $$ & 5131 (25.54 \%) & $$ & 
			4440
			(22.10 \%)  &  $$ & -\\
 \hline
  \label{table:parameters4}
  \end{tabular}
  \begin{center}
  \caption{Daily hemispheric sunspot numbers.
  Columns, rows and numbers in
  the rows have the same meaning as in Table 1.}
  \end{center}
  \end{table*}

\subsection{Results for non-integer and dimensional time series}
\begin{figure}[h]
       \centering{
      \epsfig{file=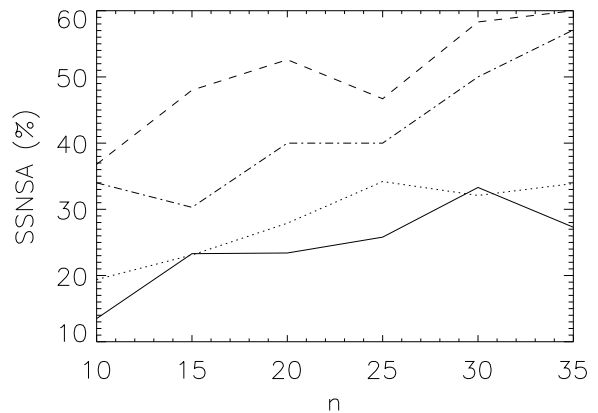, width = 9cm}}
                  \caption{Student's t-test: SSNSA of sunspot area (solid line), $H_{\alpha}$ flare
		  index (dotted line), Mount Wilson total magnetic flux
		  (dash - dot line), and Kitt Peak averaged magnetic flux density
		  (dashed line),
		  versus the number (n) of elements in each
		  group.}             
   \label{sig3} 
          \end{figure}

Here, the Student's t-test described in section~\ref{se1} has been
applied. We have chosen $\alpha = 0.01$ and $n = 30$, which means 
that for sunspot areas we have considerered groups of $30$
months; for the $H_{\alpha}$ flare index, groups of $30$ days; and for
the Mount Wilson total magnetic flux and the Kitt Peak averaged
magnetic flux density, groups of $30$ Carrigton rotations.
Table~6 shows the results for the SSNSA and in the case of sunspot areas, only $14 \%$ of the thirty-
month groups are significant at the $99 \%$ level; in the case of the $H_{\alpha}$ flare
index only $32 \%$ of the thirty-day 
groups are significant at the $99 \%$ level; in the case of the Mount Wilson total magnetic flux 
only $43 \%$ of the thirty Carrington rotation  
groups are significant at the $99 \%$ level; and in the case of
the Kitt Peak averaged magnetic flux density only $48 \%$ of the thirty
Carrington rotations groups are significant at the $99 \%$ level.
However, an important point to be considered when applying this test
is the value chosen for $n$.  We can highlight this point by
considering different values for $n$, and repeating the above
calculations.  In Fig.~\ref{sig3} the SSNSA versus $n$, number of
elements in each group, has been
plotted, and it can be seen that the significance increases with $n$.
Thus the choice of the value of $n$ is
important because it determines the significance of the asymmetry,
modifying it when $n$ is modified.

 \begin{table*}
  \small
  \centering
  \begin{tabular}{cccccccc}
  \hline
  \hline
  Sunspot areas &  \ \ \ \ \ \ & $H_{\alpha}$ Flare Index & \ \ \ \ \ \  
  & MW magnetic flux & 
  \ \ \ \ \ \  & KP magnetic flux &  \\
  \hline
  \hline
                       33 \% & $$ & 32 \%  & $$ &
		       50 \% & $$ & 58 \% \\

 \hline
  \label{table:parameters8}
  \end{tabular}
  \begin{center}
  \caption{Student's t-test. The numbers in the row define the
  percentage of
  groups of each time series that are statistically highly
  significant ($t \geq 0.99$). The value of $n$ in the Student's t-test 
  formula is $30$. }
  \end{center}
  \end{table*}

Another interesting situation is the case of solar
activity time series showing consecutive episodes with alternate
dominance between hemispheres.  Assume, for instance, that we have a
time series, corresponding to the Northern hemisphere, such as: 
\begin{eqnarray}
    North &=& 60, 60, 60, 60, 60, 60, 2, 3, 4, 5, 6, 7, 60, 60, 60, 60,
    60,  \nonumber \\
          & & 60, 2, 3, 4, 5, 6, 7, 60, 60, 60, 60, 60, 60, 2, 3, 4, 5, 6, 7 \nonumber
\end{eqnarray}
and another one, corresponding to the Southern hemisphere, such as:
\begin{eqnarray}
South &=& 1, 2, 3, 4, 5, 6, 60, 60, 60, 60, 60, 60, 1, 2, 3, 4, 5, 6, 
60, \nonumber \\
          & & 60, 60,60, 60, 60, 1, 2, 3, 4, 5, 6, 60, 60, 60, 60, 60, 60
	  \nonumber
\end{eqnarray}
representing an extreme case of consecutive strong asymmetry between
hemispheres.  First, we take $n = 6$ and determine the SSNSA between
time series, obtaining that it is highly significant ($t \geq
0.99$) for the six groups.  Next, we consider $n = 12$ and we
determine, again, the SSNSA, obtaining that the asymmetry in this case
is not significant at all ($t < 0.9$) for the 3 groups generated.
This example points out the effect of the chosen $n$ since in the
case $n = 6$ the test detects the asymmetry between hemispheres,
however, in the case $n = 12$ the test is unable to detect the
asymmetry since in this case the numerator of the statistic $\hat t$
is almost zero.  Then, in extreme cases, such as the one
illustrated above, a null result can be misleading since it could mean
that the value of $n$ corresponds to a dominance of the Northern
hemisphere followed by a dominance of the Southern hemisphere, and so on.
One way to avoid this problem is to consider $\vert D_{i}
\vert$ instead of $D_{i}$ in the numerator of the statistic $\hat t$.
Then, using this consideration in the calculations of the SSNSA of the
above synthetic series with $n = 12$, we obtain that all the groups
are highly significant ($t \geq 0.99$), as one expects due to the
behaviour of the asymmetry between hemispheres.

\subsection{Dependence of the SSNSA on the solar cycle phase}

The results shown in Tables 3-7 correspond to the SSNSA of solar
activity time series spanning different time intervals that
cover different phases of the solar activity cycle.  Due
to the difference in the covered time intervals, these time series
only overlap during a few solar cycles.  In order to study the
behaviour of the SSNSA with the phase of the solar cycle we have
considered the Northern and Southern daily number of X-ray flares and the
Northern and Southern daily hemispheric sunspot number time series.  The
first time series covers three solar cycles and the second covers
six solar cycles, and they overlap during the period $1976-2004$.  We
have split each time series as many times as possible using the
following criterion: From the middle of the descending phase of one
solar cycle to the middle of the ascending phase of the following
solar cycle, covering the minimum of solar activity, and from the
middle of the ascending phase to the middle of the descending phase of
the same solar cycle, covering the maximum of solar activity.  In this
way, we obtained $5$ and $11$ shorter time series for the daily
number of Northern and Southern X-ray flares and daily hemispheric sunspot
numbers, respectively.  Then, we applied to both time series the
statistical tests of section~\ref{se1}, except the binomial
distribution test, because the daily hemispheric sunspot numbers is a
non-integer time series.  The results obtained applying the Excess
test to the daily hemispheric sunspot number time series are shown in
Figure~4. This figure shows the differences between the SSNSA
around the maximum and the minimum of solar activity and, although they
are not large, a systematic difference appears: the SSNSA is always
higher around the minimum of solar activity.  It has been
pointed out (Swinson et al.  1986; Vizoso and Ballester, 1990) that
the North-South asymmetry of solar activity reaches very high values
around the minimum of solar activity, thus, if only a few sunspot
groups appear on the Sun and all them are in the same hemisphere, the
value of the asymmetry and the SSNSA would be very high. However, around
the maximum of solar activity, sunspot groups are more evenly
distributed between hemispheres giving place to a lower asymmetry and
SSNSA. This could be an explanation for the dependence of the
SSNSA on the solar cycle phase.

A similar behaviour appears for the case of the daily number of X-ray
flare time series, although in this case the SSNSA in any considered
phase of the solar cycle is very low, such as can be expected due to
the low SSNSA obtained when this time series is considered as a whole
(see Table 5).  The results obtained with the rest of tests are
similar to those shown in Figure~4.

    \begin{figure}[h]
       \centering{
      \epsfig{file=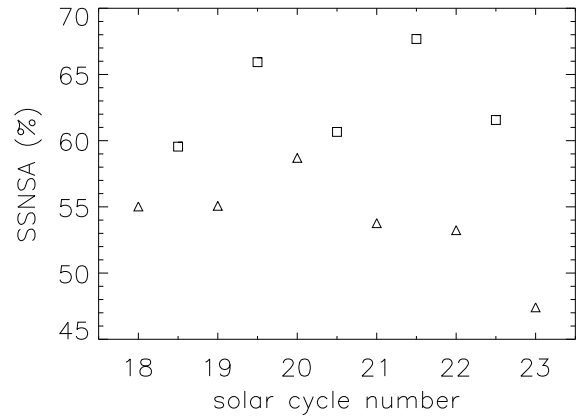, width = 9cm}}
                  \caption{SSNSA of the daily hemispheric sunspot numbers
		  versus solar cycle. Triangles denote the SSNSA
		  around the maximum of solar activity. Squares denote
		  the SSNSA around the minimum of solar activity.}             
   \label{sig4} 
          \end{figure}
 
\section{Conclusions}

We have discussed the difficulties encountered
when trying to assess the SSNSA of the most common solar activity time
series.  We have found that in the case of integer or non-integer and dimensionless
data sets several statistical tests such as the binomial distribution, normal
approximation to the binomial distribution, chi-square test and Excess
can be used, however, the obtained results strongly depend on the data
binning applied.  On the other hand, when non-integer and
dimensional data are considered, a statistical test independent of the
units can be used, the Student's t-test, but the obtained results
depend again on the value chosen for the binning.
Our results also suggest that there is a systematic 
difference between the values of the SSNSA around the maximum and the 
minimum of solar activity, which suggests that there is a
dependence of the SSNSA on the solar cycle phase.

Taking into account these results, how can we assess the SSNSA? It
seems that a definitive answer cannot be given because in one case it
strongly depends on the data binning performed using, mostly, our
terrestrial calendar (days, months, Carrington rotation, years, solar
cycles, etc.)  while solar activity does not care about it, and in the
other case the answer depends on the number of elements in each
considered group.  In the case of integer or non-integer and
dimensionless time series, the length of the bin should help to reveal
physically meaningful results.  Thus, for a dataset of daily values to
take $n=30$ would be appropiate since this would correspond to about
one solar rotation.  Furthermore, if one wants to obtain some
information about phases of solar activity such as the ascending or
the descending branch of the solar cycle, or about the period around
the maximum or the minimum of solar activity, then the length of the
bin has to be chosen in agreement with the time intervals under study.
These considerations could also be applied to the case of non-integer
and dimensional time series because when using the Student's t-test we
also need to make a choice for the value of $n$, the number of
elements in each of the groups in which we split the dataset.  Thus, a
visual inspection of time series is not appropriate to ascertain the
N-S asymmetry of solar activity time series but, on the other hand, to
determine an absolute value of the SSNSA is difficult, worsened by the
fact that the records of solar activity are not independent.

All the results obtained up to now on the SSNSA of different solar
activity time series by different authors must be considered with
care, as must be the studies performed of the behaviour of the N-S
asymmetry of different solar activity features which assume that there
is a real and significant asymmetry between hemispheres.

\section{Acknowledgements}

We acknowledge the National Geophysical Data Center, from
whose ftp server the Kandilli flare index and international sunspot number
were downloaded.  We also acknowledge the Solar Influences Data Center
(SIDC) for the compilation of the International Sunspot Number.
NSO/Kitt Peak data used here were produced cooperatively by NSF/NOAO,
NASA/GSFC, and NOAA/SEL. This study includes data about magnetic flux
from the synoptic program at the 150-Foot Solar Tower of the Mt.
Wilson Observatory and were kindly provided by J. Boyden.  The Mt.
Wilson 150-Foot Solar Tower is operated by UCLA, with funding from
NASA, ONR, and NSF, under agreement with the Mt.  Wilson Institute.
The sunspot area data were compiled by D. Hathaway of NASA's Marshall
Space Flight Center.  J. Terradas thanks the Spanish Ministry of
Science and Education for the funding provided by a Juan de la
Cierva fellowship.

\end{document}